\documentclass[prd,showpacs,preprintnumbers,amsmath,amssymb]{revtex4}

\usepackage{graphicx}
\usepackage{dcolumn}
\usepackage{bm}

\begin{document}

\title{Vacuum Polarization and Persistence on the Black Hole Horizon}

\author{Sang Pyo Kim}\email{sangkim@kunsan.ac.kr}
\affiliation{Department of Physics, Kunsan National University,
Kunsan 573-701, Korea}
\affiliation{Institute of Astrophysics, Center for Theoretical Physics, Department of Physics, National Taiwan University,
Taipei 106, Taiwan}

\author{W-Y. Pauchy Hwang}\email{wyhwang@phys.ntu.edu.tw}
\affiliation{APCosPA and LeCosPA, Institute of Astrophysics, Center for Theoretical Physics, Department of Physics, National Taiwan University, Taipei 106, Taiwan}

\date{\today}

\begin{abstract}
  We find the exact one-loop effective action in a Schwarzschild black hole in the proper-time integral from the Schwinger variational principle, which has the same form up to number of states as the Heisenberg-Euler and Schwinger QED effective action in a constant electric field. The leading term for Hawking radiation comes from the first simple pole of the vacuum polarization and the sum of residues from all simple poles exactly leads to the vacuum persistence.
 We show that the vacuum persistence is the total flux of Hawking radiation for bosons and fermions and the vacuum polarization has an expression of thermal distribution and the propagator.
\end{abstract}

\maketitle

{\it Introduction} $-$ A curved spacetime and an external gauge field can produce particle pairs from vacuum fluctuations. Hawking radiation from black holes \cite{Hawking} and Schwinger pair production from electric fields \cite{Schwinger} are the most recognized phenomena in theoretical physics. In the tunneling picture for Hawking radiation one species of particle pairs produced from vacuum fluctuations near the black hole horizon is emitted to spatial infinity while the other species falls into the black hole \cite{Parikh-Wilzcek}. Hawking radiation of bosons or fermions obeys the Bose-Einstein (BE) or Fermi-Dirac (FD) distribution \cite{Hawking}. In quantum electrodynamics (QED) a strong electric field whose potential energy of a charged particle across the Compton wavelength is comparable to the rest mass energy of the pair separates the virtual pairs into real ones.

Particle production is one of nonperturbative aspects of quantum field theory in background fields. The nonperturbative effective action in QED has been known for many decades since quantum field theory was formulated. Heisenberg and Euler used the electron-positron theory to obtain the effective action in a constant electromagnetic field, whose imaginary part is responsible for pair production from the Dirac sea and the decay of the vacuum \cite{Heisenberg-Euler}. Schwinger obtained the nonperturbative one-loop effective action in the proper-time integral \cite{Schwinger}. The effective action is characterized by the vacuum persistence, twice of the imaginary part, and the vacuum polarization, the real part. The vacuum persistence is entirely determined by the mean number of produced pairs. In spite of intensive studies for Hawking radiation, the effective action for black holes in the sense of QED action by Heisenberg-Euler and Schwinger has not been found yet.

In this Letter we find the exact one-loop effective action for Hawking radiation in a Schwarzschild black hole in a complete analogy with the Heisenber-Euler and Schwinger effective action in a constant electric field. To do that, we use the effective action from the Schwinger variational principle \cite{Schwinger51} in the in-/out-state formalism developed by DeWitt \cite{DeWitt}. In this formalism pair production from external background fields, which may restrict the consistency of quantum field theory \cite{Gavrilov-Gitman}, can be incorporated. The vacuum persistence is determined by the mean number of produced pairs and vice versa, which imposes a consistency condition as in QED. Furthermore we show that the vacuum persistence is the total flux of Hawking radiation from the gravitational anomalies \cite{Robinson-Wilczek,IUW} and that the vacuum polarization can be written as a thermal effective action with the Hawking temperature. We also find the effect of the amplification or gray body factor on the effective action, introduce a renormalization scheme for the effective action, and discuss the vacuum persistence for a charged rotating black hole.

{\it Vacuum Persistence Amplitude} $-$ Particle production from an external gauge field or a curved spacetime can be formulated through the Bogoliubov transformations [in Planck units of $c = \hbar = G= k_{B} =1$]
\begin{eqnarray}
a_{J, {\rm out}} = \alpha_{J} a_{J, {\rm in}} + \beta_{J} b^{\dagger}_{J, {\rm in}},
\end{eqnarray}
where $a_J$ is the annihilation operator for particle, $b^{\dagger}_J$ is the creation operator for antiparticle, and $J$ collectively denotes all the quantum number for each mode. The coefficients $\alpha$ and $\beta$ become matrices in the case of mode-mixing. The in-vacuum is defined by $a_{J, {\rm in}} \vert 0, {\rm in} \rangle = 0$ and the out-vacuum by $a_{J, {\rm out}} \vert 0, {\rm out} \rangle = 0$ for all $J$. The spin statistics theorem on the quantization of fields requires the Bogoliubov relation for each $J$
\begin{eqnarray}
|\alpha_{J}|^2 \mp |\beta_{J}|^2 = 1, \label{bog rel}
\end{eqnarray}
where the upper sign is for bosons and the lower sign for fermions and this convention will be used hereafter.
The mean number of produced pairs with $J$ is given by
\begin{eqnarray}
N_{J} = \langle 0, {\rm out} \vert a^{\dagger}_{J, {\rm in}} a_{J, {\rm in}} \vert 0, {\rm out} \rangle = |\beta_{J}|^2.
\end{eqnarray}

The Schwinger variational principle gives the vacuum persistence amplitude, the action for the effective Lagrangian density, in the in-/out-state formalism as \cite{Schwinger51}
\begin{eqnarray}
e^{i W} = \langle 0, {\rm out} \vert 0, {\rm in} \rangle,
\end{eqnarray}
and therefrom follows the effective action (Lagrangian) density in a curved spacetime \cite{DeWitt}
\begin{eqnarray}
W = \int d^4x \sqrt{-g} {\cal L}_{\rm eff} (x) =\mp i \sum_{J} \int \ln (\alpha^*_{J}). \label{eff act}
\end{eqnarray}
In the above the summation-integral $\sum_J \int$ will be taken over all number of states and over the spacetime region where particles are effectively produced.
Then the Bogoliubov relation (\ref{bog rel}) leads to the vacuum persistence
\begin{eqnarray}
2 {\rm Im} (W) =  \pm \sum_{J} \int \ln ( 1 \pm N_{J}). \label{vac per}
\end{eqnarray}
The vacuum persistence and the mean number of produced pairs should always satisfy the relation (\ref{vac per}) as a consistency condition just as the Heisenberg-Euler and Schwinger effective action does in QED. An interesting observation made by Stephens \cite{Stephens} is that the vacuum persistence for the BE- and FD-distribution now takes the Nikishov's representation in QED \cite{Nikishov} with the opposite sign
\begin{eqnarray}
2 {\rm Im} (W) = \mp  \sum_J \int \ln ( 1 \mp e^{- \beta \omega_J}).
\end{eqnarray}
That $2 {\rm Im} (W)$ is equal to the logarithm of partition function, $\ln Z$, plus the vacuum energy allows one to interpret the vacuum persistence per volume as the pressure for boson and fermion gas \cite{Ritus}.

{\it Vacuum Polarization on the Horizon} $-$ We now study the effective action (\ref{eff act}) for black holes and the vacuum persistence (\ref{vac per}) for Hawking radiation. For simplicity we consider a Schwarzschild black hole with the metric
\begin{eqnarray}
ds^2 = - f(r) dt^2 + \frac{dr^2}{f(r)} + r^2 d \Omega_{2}, \quad f(r) = 1 - \frac{2M}{r}.
\end{eqnarray}
Hawking radiation has the BE- or the FD-distribution with the Hawking temperature $T_{\rm H} = \kappa/2 \pi$ for a given quantum $J$ (for a review and references, see \cite{Page-review})
\begin{eqnarray}
N_{J} = \frac{1 - |R_J|^2}{e^{\beta \omega} \mp 1}, \label{haw rad}
\end{eqnarray}
where $\kappa = f'(r_H)/2$ is the surface gravity on the horizon $r_H = 2M$, $\beta = 1/T_{\rm H}$ is the inverse of thermal energy, and the reflection coefficient $|R_J|^2$ for the incident wave $J$ is an amplification factor.

The Bogoliubov coefficients for emission of massless bosons with energy $\omega$ are given by \cite{DeWitt}
\begin{eqnarray}
\alpha_{J} &=& A_{J} e^{\pi \omega/2 \kappa} \Gamma (1 + i \frac{\omega}{\kappa}), \nonumber\\
\beta_{J} &=& - A_{J} e^{- \pi \omega/2 \kappa} \Gamma (1 + i \frac{\omega}{\kappa}). \label{bog coef}
\end{eqnarray}
The factor $A_{J}$ will not be specified since it is not necessary for the renormalized effective action.
Hawking radiation from coefficients (\ref{bog coef}) is given by the BE-distribution (\ref{haw rad}) with $R_J =0$. The effective action (\ref{eff act}) for bosons, after deleting any terms that will be removed through renormalization, is then given by
\begin{eqnarray}
W = i \sum_{J} \int \ln \Bigl( \Gamma (1 - i \frac{\beta \omega}{2 \pi}) \Bigr).
\end{eqnarray}
The summation-integral for emission of particles with mass $\mu$ is the number of states on the horizon of the black hole and is explicitly given by
\begin{eqnarray}
\sum_J \int = \frac{4 \pi r_H^2}{\beta} \sum_{l, m, p} \int_{\mu}^{\infty} \frac{d \omega}{2 \pi},
\end{eqnarray}
where $l$ is the angular number for spheroidal harmonics, $m$ is the axial angular momentum, and $p$ is the polarization of the mode. As shown in Table I, the factor $1/\beta$ corresponds to the number of states $qE/(2 \pi)$ in the Heisenberg-Euler and Schwinger effective action \cite{KLY}, which can be interpreted as the Davies-Unruh temperature times the Compton wavelength of the charge $q$ \cite{Hwang-Kim}.

The integral representation for the Gamma-function \cite{gamma-int}
and the Cauchy residue theorem along an infinite quarterly circle in the first quadrant \cite{KLY,Kim10} lead to
\begin{eqnarray}
\int_0^{\infty} \frac{dz}{z} \frac{e^{- (1 - i \beta \omega/(2 \pi)) z}}{1 - e^{-z}} = {\cal P} \int_0^{\infty} \frac{ds}{s} \frac{e^{- (1 - i \beta \omega/(2 \pi)) (is)}}{1 - e^{- i s}}
 + \pi i \sum_{n =1}^{\infty}\frac{e^{- \beta \omega n}}{i (2 n \pi)}, \label{gam-reg}
\end{eqnarray}
where ${\cal P}$ is the principal value. We thus obtain the reduced effective action in the $(t-r)$ plane for massless bosons
\begin{eqnarray}
{\cal L}^{\rm bos}_{\rm red} = - \frac{1}{2 \beta} \sum_{l, m, p}  \int_0^{\infty} \frac{d \omega}{2 \pi} \int_0^{\infty} \frac{ds}{s} e^{- \beta \omega s/(2 \pi)} \frac{\cos(\frac{s}{2})}{\sin (\frac{s}{2})} - \frac{i}{2 \beta} \sum_{l, m, p} \int_0^{\infty} \frac{d \omega}{2 \pi} \ln (1 - e^{- \beta \omega}), \label{vac-bos}
\end{eqnarray}
and, similarly, for massless fermions
\begin{eqnarray}
 {\cal L}^{\rm fer}_{\rm red} =  \frac{1}{2 \beta} \sum_{l, m, p} \int_0^{\infty} \frac{d \omega}{2 \pi} \int_0^{\infty} \frac{ds}{s} e^{- \beta \omega s/(2 \pi)} \frac{1}{\sin (\frac{s}{2})} + \frac{i}{2 \beta} \sum_{l, m, p} \int_0^{\infty} \frac{d \omega}{2 \pi} \ln (1 + e^{- \beta \omega}). \label{vac-fer}
\end{eqnarray}
Then the vacuum persistence $2 {\rm Im} ({\cal L}_{\rm red})$ in Eqs. (\ref{vac-bos}) and (\ref{vac-fer}) are consistent with the relation (\ref{vac per}). Remarkably, as shown in Table I, the vacuum polarization and persistence (\ref{vac-bos}) for bosons has the same form as that for spinor QED while the vacuum polarization and persistence (\ref{vac-fer}) for fermions takes the same form as that for scalar QED \cite{Schwinger,KLY}. This apparently contradicts the spin statistics theorem for QED. However, this paradox may be resolved by the observation by M\"{u}ller, Greiner and Rafelski that the vacuum polarization of QED can be written with the right spin statistics \cite{MGR} and by another observation by us that the spin statistics for the vacuum persistence can also be inverted \cite{Hwang-Kim}.

\begin{table}
\caption{\label{Tab1} Comparison between Schwarzschild Black Hole and QED}
\begin{ruledtabular}
 \begin{tabular}{lll}
 {\bf Notation} & {\bf Schwarzschild Black Hole} & {\bf QED}\\
 \hline
$\frac{1}{\beta} = k_B T$ & $\frac{\kappa}{2 \pi}$ & $\frac{(qE/m)}{2 \pi}$\\
Number of states $\sum_J \int$  & $\frac{1}{\beta} \sum_{l, m, p} \int \frac{d\omega}{2 \pi}$ & $\frac{m}{\beta}
\sum_{\sigma} \int \frac{d^2 {\bf k}_{\perp}}{(2\pi)^2}$ \\
Vacuum persistence\footnote{The upper sign is for bosons and the lower sign for fermions.} & $\mp (\sum_J \int) \ln (1 \mp e^{- \beta \omega})$ & $\pm (\sum_J \int) \ln (1 \pm e^{- \beta (\frac{{\bf k}^2_{\perp}}{2m} + \frac{m}{2})})$ \\
Vacuum polarization\footnote{In the black hole $\{ \cos(\frac{s}{2}) \}$ appears only for bosons and equals to unity for fermions, whereas in QED $[\cos(\frac{s}{2})]$ appears only for fermions and equals to unity for bosons.} & $\mp \frac{1}{2} (\sum_J \int) \int_{0}^{\infty} \frac{ds}{s} e^{- \beta \omega s/(2 \pi)} \frac{\{ \cos(\frac{s}{2}) \}}{\sin (\frac{s}{2})} $ & $\pm (\sum_J \int) \int_{0}^{\infty} \frac{ds}{s} e^{- \beta (\frac{{\bf k}^2_{\perp}}{2m} + \frac{m}{2})s/(2 \pi)} \frac{[\cos(\frac{s}{2})]}{\sin (\frac{s}{2})}$
 \end{tabular}
\end{ruledtabular}
\end{table}

We advance a renormalization scheme for the vacuum polarization (\ref{vac-bos}) and (\ref{vac-fer}). For particles with mass $\mu$ as the IR-cutoff, after the integration $\int_{\mu}^{\infty} d \omega/2\pi$, we get
\begin{eqnarray}
{\rm Re} {\cal L}^{\rm ren}_{\rm red} = \mp \frac{1}{2 \beta^2} \sum_{l, m, p}  \int_0^{\infty} \frac{ds}{s^2} e^{- \beta \mu s} \Bigl( \frac{ \{ \cos(\frac{s}{2}) \}}{\sin (\frac{s}{2})} - \frac{a_0}{s} - a_1 s \Bigr),
 \label{vac-ren}
\end{eqnarray}
where $\{ \cos(s/2) \}$ appears only for boson emission and equals to unity for fermion emission. Here we used the Schwinger's prescription \cite{Schwinger}, which subtracts all the divergent terms: $a_0 = 2, a_1 = 1/12$ for bosons and $a_0 = 2, a_1 = -1/6$ for fermions. The $a_0$-term leads to $\Gamma(-2) \mu^2$ and the $a_1$-term leads to $\Gamma (0)/\beta^2$, which after dividing by the factor of $\beta^2/4 \pi$ from the area of the horizon will renormalize the gravitational constant and the coupling constant for one-loop terms. For a large black hole mass $(\beta \gg 1)$, the vacuum polarization has a power series of $1/(\beta^2 (\beta \mu)^{2(n-1)})$ for $n = 2, 3, \cdots$ from expansion of $\{ \cos(s/2) \}/\sin(s/2)$, whereas for a small black hole $( \beta < 1)$, it has the leading behavior $(1/\beta^2) \ln (1/\beta \mu)$, in which case higher loops should be included.
The vacuum persistence has already been renormalized since it is the sum of the residues of simple poles for the renormalized effective action.

{\it Vacuum Persistence and Gravitational Anomalies} $-$  A physical interpretation of the vacuum persistence is the decay rate of the vacuum due to Hawking radiation. In fact, the effective action for any quantum system suffering instability from pair production in background fields takes a complex value, whose vacuum persistence determines the decay rate by (\ref{vac per}). It has been known for many years that Hawking radiation and Schwinger mechanism can be interpreted as the trace anomalies \cite{Christensen-Fulling,Dittrich-Sieber}. Recently the flux of Hawking radiation is shown to be a consequence of the gravitational anomalies in black hole spacetimes \cite{Robinson-Wilczek,IUW}. Since the trace and the gravitational anomalies are quantum effects in background fields and are related with pair production, one may expect that the vacuum persistence of the black hole, the one-loop result, should have a connection with Hawking radiation. Directly integrating Eqs. ({\ref{vac-bos}) and (\ref{vac-fer})
\begin{eqnarray}
2 {\rm Im} ({\cal L}^{\rm bos}_{\rm red}) =  \sum_{l, m, p} \frac{\pi}{12} \frac{1}{\beta^2}, \quad
2 {\rm Im} ({\cal L}^{\rm fer}_{\rm red}) =  \sum_{l, m, p} \frac{\pi}{24} \frac{1}{\beta^2},
\end{eqnarray}
the vacuum persistence is the total flux of Hawking radiation from the gravitational anomalies \cite{Robinson-Wilczek}. Note that the renormalized value of two-point function on the horizon is $\langle \phi^2 (x) \rangle_{\rm HH}^{\rm ren} =1/3 \beta^2$ in the Hartle-Hawking vacuum \cite{Candelas}.

{\it Thermal Interpretation of Vacuum Polarization} $-$ We want to show that the vacuum polarization (\ref{vac-bos}) and (\ref{vac-fer}) can be written in the form of a thermal effective action with the right spin statistics. Following Ref. \cite{MGR}, we use the expansion of
${\rm cotan} (x) - x$ and ${\rm cosec} (x) - x$ \cite{math}, rescale the integration variable and sum over $n$, to write the reduced effective action for bosons or fermions after renormalization as
\begin{eqnarray}
 {\cal L}_{\rm red} = \pm \frac{1}{\pi \beta} \sum_{l, m, p}  \int_0^{\infty} \frac{d \omega}{2 \pi} \int_0^{\infty} ds \frac{\omega}{s^2 - \omega^2} \ln (1 \mp e^{- \beta s}).
 \label{ther-vac}
\end{eqnarray}
The principal value of (\ref{ther-vac}) correctly yields the vacuum persistence in Eqs. (\ref{vac-bos}) and (\ref{vac-fer}). Integrating by parts with respect to $s$, we rewrite the vacuum polarization as a thermal effective action,
\begin{eqnarray}
 {\cal L}_{\rm red} = \pm  \frac{1}{\pi} \sum_{l, m, p}  \int_0^{\infty} \frac{d \omega}{2 \pi} \int_0^{\infty} ds  \frac{s} {e^{\beta s} \mp 1} \int_0^{\omega} \frac{d \omega'}{s^2 - \omega'^2}. \label{ther-eff-bos}
\end{eqnarray}
The vacuum polarization has the form of the mean thermal energy either in the BE- or in the FD-distribution times the integral of the propagator. The effective action is taken in a two-dimensional Minkowski spacetime since Hawking radiation of the Schwarzschild black hole takes place along the radial direction, thus being effectively two dimensional. It should be noted that the effective action (\ref{ther-eff-bos}) is not renormalized since we have subtracted only the most singular term in Eq. (\ref{vac-ren}).

{\it Discussions} $-$ First, we consider the effect of the amplification or gray body factor  on the effective action since it modifies Hawking radiation from the BE- or FD-distribution.
Then the vacuum persistence from Eq. (\ref{vac per}) takes the form
\begin{eqnarray}
2 {\rm Im} ({\cal L}^{\rm full}_{\rm red}) =  \mp \frac{1}{\beta} \sum_{l, m, p} \int_0^{\infty} \frac{d \omega}{2 \pi}  \ln \Bigl( \frac{1 \mp e^{- \beta \omega}}{1 \mp
|R_J|^2 e^{- \beta \omega}} \Bigr),
\end{eqnarray}
and the corresponding vacuum polarization takes the form
\begin{eqnarray}
{\rm Re} ({\cal L}^{\rm full}_{\rm red}) = \mp \frac{1}{2 \beta} \sum_{l, m, p}  \int_0^{\infty} \frac{d \omega}{2 \pi} \int_0^{\infty} \frac{ds}{s} e^{- \beta \omega s/(2 \pi)} \Bigl(\frac{ \{ \cos(\frac{s}{2}) \}}{\sin (\frac{s}{2})} (1 - e^{2 \ln (|R_J|) s}) - \cdots \Bigr).
\end{eqnarray}

Second, since the formula (\ref{vac per}) is exact for Hawking radiation, the vacuum persistence can be investigated for a charged rotating black hole
\begin{eqnarray}
2 {\rm Im} (W) = \mp \sum_J \ln \Bigl( 1 \mp e^{- \beta (\omega - m \Omega_{\rm H} - q \Phi_{\rm H})} \Bigr), \label{rot bh}
\end{eqnarray}
where $\Omega_{\rm H}$  and $\Phi_{\rm H} = Q/r_{\rm H}$ are the angular velocity and electric potential of the black hole, and $m$ and $q$ are the axial angular momentum and charge of emitted particles. As the vacuum decays more efficiently through the channel with the largest ${\rm Im} (W)$ than channels with small ${\rm Im} (W)$, the black hole emits more rapidly charges of the same kind as the black hole than charges of the opposite kind and so does the angular momentum, as shown by Page \cite{Page}. For the superradiant mode of bosons the absolute value will be taken inside the logarithm function.

Third, though the four-dimensional Schwarzschild black hole is considered, the result of this paper that is expressed in terms of the surface gravity can be extended to higher dimensions since Hawking radiation takes place in the $(t-r)$ plane. The number of states should properly include the higher spheroidal harmonics and the area of the horizon. The vacuum persistence (\ref{rot bh}) for the charged rotating black hole may lead to the vacuum polarization in a similar form as (\ref{vac-bos}) or (\ref{vac-fer}), which, however, has to be directly derived from the Bogoliubov coefficients.

{\it Summary} $-$ We have obtained the exact one-loop effective action for Hawking radiation in a Schwarzschild black hole. As summarized in Table I, the effective action in black hole has the same form as the Heisenberg-Euler and Schwinger QED effective action in a constant electric field, except that Hawking radiation for bosons corresponds to spinor QED and that for fermions to scalar QED. The vacuum persistence, twice of the imaginary part, equals to the total flux of Hawking radiation from the gravitational anomalies. We show that the vacuum polarization has an expression of thermal effective action with the correct spin statistics in a two-dimensional Minkowski spacetime. The effect of the amplification or gray body factor on the effective action, the renormalization scheme by Schwinger and the vacuum persistence of a charged rotating black hole are discussed.

\acknowledgments

S.~P.~K. would like to thank Andreas Wipf for helpful discussions.
The work of S.~P.~K.~was supported in part by the National Research Foundation (NRF) Grant funded by the
Korean Government (MEST)(2010-0016-422). The work of W-Y.~P.~H.~was supported in part by National Science Council Project (NSC 99-2112-M-002-009-MY3). The visit of S.~P.~K. to the Institute of Astrophysics, National Taiwan University, was supported by National Science Council Grant (NSC 100-2811-M-002-012).

\end{document}